\documentclass[final,authoryear,5p]{elsarticle}

\usepackage{epsfig}

\usepackage{amssymb}

\usepackage[ps2pdf,%
a4paper=true,%
breaklinks=true,%
colorlinks=true,%
pdfauthor={First Author et al.},%
pdftitle={Template for manuscripts in Advances in Space Research}%
]{hyperref}

\journal{Advances in Space Research}

\begin{document}


\begin{frontmatter}

\title{Quantum Stark broadening of Ar XV lines. Strong collision and quadrupolar potential contributions}

\author[label1,label2]{H. Elabidi\corref{cor}}
\address[label1]{Deanship of the Foundation Year, Umm Al-Qura University, Makkah, KSA}
\address[label2]{GRePAA, Facult\'{e} des Sciences de Bizerte,
Universit\'{e} de Carthage, Tunisia} \cortext[cor]{Corresponding
author}\ead{haelabidi@uqu.edu.sa, haykel.elabidi@fsb.rnu.tn}

\author[label3]{S. Sahal-Br\'{e}chot}
\address[label3]{LERMA, Obs. Paris, UMR CNRS 8112,
UPMC, B\^{a}timent Evry Schatzman, France}
 \ead{sylvie.sahal-brechot@obspm.fr}

\author[label3,label4]{M. S. Dimitrijevi\'{c}}
\address[label4]{Astronomical Observatory, Volgina 7, 11060
Belgrade 38, Serbia}  \ead{mdimitrijevic@aob.rs}

\begin{abstract}
We present in this paper electron impact broadening for six Ar XV
lines using our quantum mechanical formalism and the semiclassical
perturbation one. Additionally, our calculations of the
corresponding atomic structure data (energy levels and oscillator
strengths) and collision strengths are given as well. The lines
considered here are divided into two sets: a first set of four
lines involving the ground level: 1s$^{2}2$s$^{2}$ $^{1}$S$_{0}-$
1s$^{2}2$s$n$p $^{1}$P$^{\rm o}_{1}$ where $2\leq n \leq5$ and a
second set of two lines involving excited levels: 1s$^{2}2$s2p
$^{1}$P$^{\rm o}_{1}-$1s$^{2}2$s3s $^{1}$S$_{0}$ and 1s$^{2}2$s2p
$^{3}$P$^{\rm o}_{0}-$1s$^{2}2$s3s $^{3}$S$_{1}$. An extensive
comparison between the quantum and the semiclassical results was
performed in order to analyze the reason for differences between
quantum and semiclassical results up to the factor of two. It has
been shown that the difference between the two results may be due
to the evaluation of strong collision contributions by the
semiclassical formalism. Except few semiclassical results, the
present results are the first to be published. After the recent
discovery of the far UV lines of Ar VII in the spectra of very hot
central stars of planetary nebulae and white dwarfs, the present
-and may be further- results can be used also for the
corresponding future spectral analysis.

\end{abstract}


\end{frontmatter}

\parindent=0.5 cm

\section{Introduction}

The Stark broadening mechanism is important in stellar
spectroscopy and in the analysis of astrophysical and laboratory
plasmas. Its influence should be considered for the opacity
calculations, the modelling of stellar interiors, the estimation
of radiative transfer through the stellar plasmas and for the
determination of chemical abundances of elements
\citep{Dimitrijevic03}. The need for spectral line broadening
calculations is stimulated by the development of computers.
Moreover, the development of instruments and space astronomy, such
as the new X$-$ray space telescope {\it Chandra}, stimulated the
calculations of line broadening of trace elements in the X$-$ray
wavelength range. \citet{Barstow98} have shown that analysis of
white dwarf atmospheres, where Stark broadening is dominant
compared to the thermal Doppler broadening, needs models taking
into account heavy element opacity. Consequently, atomic and line
broadening data for many elements are needed for stellar plasma
research. The recent discovery of the far UV lines of Ar VII in
the spectra of very hot central stars of planetary nebulae and
white dwarfs \citep{Werner07} showed the astrophysical interest
for atomic and line broadening data for this element in various
ionization stages. Ar XV is one of these important ions. The only
Ar XV line broadening calculations existing in the literature are
the semiclassical ones \citep{Dimitrijevic10}, where the authors
claimed that there are no experimental or other theoretical
results for a comparison.

The calculations performed in the present paper are based on the
quantum mechanical approach and the semiclassical perturbation
one. The quantum mechanical expression for electron impact
broadening calculations for intermediate coupling was obtained in
\citet{Elabidi04}. We performed the first calculations for the
$2s3s-2s3p$ transitions in Be-like ions from nitrogen to neon
\citep{Elabidi07,Elabidi08a} and for the $3s-3p$ transitions in
Li-like ions from carbon to phosphor \citep{Elabidi08b,Elabidi09}.
This approach was also used in \citet{Elabidi11a} to check the
dependence on the upper level ionization potential of electron
impact widths, and in \citet{Elabidi11b} to provide some missing
line broadening data for the C IV, N VI, O VI and F VII resonance
lines. In our quantum approach, all the parameters required for
the calculations of the line broadening such as radiative atomic
data (energy levels, oscillator strengths...) or collisional data
(collision strengths or cross sections, scattering matrices...)
are evaluated during the calculation and not taken from other data
sources. We used the sequence of the UCL atomic codes
SUPERSTRUCTURE/DW/JAJOM that have been used for many years to
provide fine structure wavefunctions, energy levels, wavelengths,
radiative probability rates and electron impact collision
strengths. Recently they have been adapted to perform line
broadening calculations \citep{Elabidi08a}. The semiclassical
perturbation formalism is described in
\citet{SSB69a,SSB69b,SSB74,Fleurier77} and updated by
\citet{DSB84,DSB95}. The atomic structure data (energy levels and
oscillator strengths) used by the semiclassical formalism for the
evaluation of line broadening are taken from the code
SUPERSTRUCTURE \citep{Eissner74}.

We will analyze here as well the reasons for discrepancies (up to
factor 2) between results for electron broadening of isolated non
hydrogenic ion lines obtained with semiclassical and quantum
methods as as were used by \citet{Ralchenko01,Ralchenko03,
Alexiou06}. For example \citet{Ralchenko01} obtained, using
quantum-mechanical method, electron-impact widths of the
2s3s$-$2s3p singlet and triplet lines of the beryllium-like ions
from B II to O V, and found that their results are generally
smaller from most semiclassical widths. In \citet{Ralchenko03},
the similar conclusion was obtained for  electron-impact widths of
the 3s$-$3p transitions in Li-like ions from B III to Ne VIII. It
was also found that  the difference between experimental and
quantum results monotonically increases with the spectroscopic
charge of an ion. \citet{Alexiou06} investigated the reasons for
discrepancies of electron-impact widths of isolated ion lines,
obtained with semiclassical non-perturbative and fully quantum
close-coupling and convergent close-coupling calculations, and
they concluded that the major reason is the neglect of penetration
by the semiclassical calculations. They also obtained and analyzed
data for Li-like 3s$-$3p from Be III to Ne VIII, Be-like  2s3s
$^{3}$S$-$2s3p$^{3}$P from C III to Ne VII and Be-like 2s3s
$^{1}$S$-$2s3p$^{1}$P from N IV to N VII. In order to contribute
to the clarification of this problem, it is of interest to compare
quantum and semiclassical results and for a more highly charged
ion like Ar XV, which is one of objectives of the present work.

In the present paper, Stark widths for six Ar XV lines will be
calculated using the two described formalisms and an extensive
comparison between the two results will be performed, in order to
contribute to the explanation of reasons for discrepancies found
in some cases, for line widths for ions in lower ionization stages
than Ar XV. It is also of interest to compare two methods for such
a higher ionization stage. Besides the Stark broadening data, we
will present the results of our calculations of the corresponding
atomic structure data (energy levels and oscillator strengths) and
collision strengths.

\section{Outline of the quantum approach and computational procedure}
We present here an outline of the quantum formalism of electron
impact broadening. More details have been given elsewhere
\citep{Elabidi04,Elabidi08a}. The calculations are made within the
framework of the impact approximation, which means that the time
interval between collisions is much longer than the duration of a
collision. The expression of the Full Width at Half Maximum $W$
obtained in \citet{Elabidi08a} is :
\begin{eqnarray} W&=&2N_{e}\left( \frac{\hbar }{m}\right) ^{2}\left(
\frac{2m\pi }{k_{B}T}\right) ^{ \frac{1}{2}}  \nonumber \\
&&\times\int\limits_{0}^{\infty }\Gamma _{w}\left(
\varepsilon\right) \exp \left( -\frac{\varepsilon}{k_{B}T}\right)
d\left( \frac{\varepsilon}{k_{B}T}\right), \label{integw}
\end{eqnarray} where $k_{B}$ is the Boltzmann constant,
$N_{e}$ is the electron density, $T$ is the electron temperature
and
\begin{eqnarray}
\Gamma _{w}(\varepsilon)
&=&\sum_{{J_{i}^{T}J_{f}^{T}lK_{i}}{K_{f}}} \frac{\left[
K_{i},K_{f},J_{i}^{T},J_{f}^{T}\right] }{2} \nonumber \\
&&\times\left\{
\begin{array}{c}
J_{i}K_{i}l \\
K_{f}J_{f}1
\end{array}
\right\} ^{2}\left\{
\begin{array}{c}
K_{i}J_{i}^{T}s \\
J_{f}^{T}K_{f}1
\end{array}
\right\} ^{2}  \nonumber \\
&&\times \left[ 1-\left( {\rm Re}\,(S_{i}){\rm Re}\,(S_{f})+{\rm
Im}\,(S_{i}){\rm Im}\,(S_{f})\right) \right], \nonumber
\\
\label{w2}
\end{eqnarray}
where \textbf{$L_{i}$}+\textbf{$S_{i}$} =\textbf{$J_{i}$},
\textbf{$J_{i}$}+\textbf{$l$} =\textbf{$K_{i}$} and
\textbf{$K_{i}$}+\textbf{$s$} =\textbf{$J_{i}^{T}$}. \textbf{$L$}
and \textbf{$S$} represent the atomic orbital angular momentum and
spin of the target, \textbf{$l$} is the electron orbital momentum,
the superscript $T$ denotes the quantum numbers of the total
electron+ion system. $S_{i}$ ($ S_{f}$) are the scattering matrix
elements for the initial (final) levels, expressed in the
intermediate coupling approximation, $\rm Re\, (S)$ and $\rm Im\,
(S)$ are respectively the real and the imaginary parts of the
S-matrix element, $\left\{
\begin{array}{c}
a b c \\
d e f
\end{array}
\right\} $ represent 6--j symbols and we adopt the notation $[x,
y, ...] = (2x + 1)(2y + 1)$... Both $S_{i}$ and $S_{f}$ are
calculated for the same incident electron energy
$\varepsilon=mv^{2}/2$. The equation (\ref{integw}) takes into
account fine structure effects and relativistic corrections
resulting from the breakdown of the $LS$ coupling approximation
for the target.

The atomic structure and the collisional data are needed for line
broadening evaluation. The atomic structure in intermediate
coupling is performed through the SUPERSTRUCTURE code (SST)
\citep{Eissner74}. The scattering problem in $LS$ coupling is
carried out by the DISTORTED WAVE (DW) code \citep{Eissner98} as
in \citet{Elabidi08a}. This weak coupling approximation for the
collision part assumed in DW is adequate for highly charged ions
colliding with electrons since the close collisions are of small
importance. The JAJOM code \citep{Saraph78} is used for the
scattering problem in intermediate coupling. {\bf R}-matrices in
intermediate coupling and real ($\rm Re\, \mathbf{S}$) and
imaginary part ($\rm Im\, \mathbf{S}$) of the scattering matrix
{\bf S} have been calculated using the transformed version of
JAJOM (Elabidi \& Dubau, unpublished results) and the program RtoS
(Dubau, unpublished results) respectively. The evaluation of $\rm
Re\, \mathbf{S}$ and $\rm Im\, \mathbf{S}$ is done according to:
\begin{eqnarray} {\rm Re}\, \mathbf{S}=\left( 1-\mathbf{R}^{2}\right)
\left( 1+\mathbf{R} ^{2}\right) ^{-1}, {\rm Im}\
\mathbf{S}=2\mathbf{R}\left( 1+\mathbf{R}^{2}\right) ^{-1}
\nonumber
\end{eqnarray}
The relation $\bf{S}=(1+i\bf{R})(1-i\bf{R})^{-1}$ guarantees the
unitarity of the {\bf S}-matrix.

\section{Results and discussion}
We present in the following subsections some atomic data and line
broadening data for Ar XV. Energy levels are compared to the
available theoretical \citep{Bhatia08,NIST12} and experimental
\citep{Edlen83,Edlen85,Khardi94,Lepson03} results. Oscillator
strengths and collision strengths for the Ar XV lines are compared
to the available theoretical results \citep{Bhatia08}. Electron
impact full widths at half maximum (FWHM) in \AA\ ($W=2w$) for the
considered Ar XV lines are calculated for a range of electron
temperatures from $5\times 10^{5}$ K to 2 $\times 10^{6}$ K and
for an electron density of 10$^{20}$ cm$^{-3}$. We choose four
lines involving the ground level (1s$^{2}2$s$^{2}$ $^{1}$S$_{0}-$
1s$^{2}2$s$n$p $^{1}$P$^{\rm o}_{1}$ where $2\leq n \leq5$) and
two others involving excited levels (1s$^{2}2$s2p $^{1}$P$^{\rm
o}_{1}-$1s$^{2}2$s3s $^{1}$S$_{0}$ and 1s$^{2}2$s2p $^{3}$P$^{\rm
o}_{0}-$1s$^{2}2$s3s $^{3}$S$_{1}$). Calculations are based on the
quantum mechanical and the semiclassical perturbation formalisms.

\subsection{Structure and electron scattering data}
The configurations used in the atomic structure description are
1s$^{2}$(2s$^{2}$, 2s2p, 2p$^{2}$, 2s$nl$) where 3$\leq n \leq 5$
and $l=$s,p,d. This set of configurations gives rise to 118 fine
structure levels. In the code SST, the wave functions are
determined by diagonalization of the non relativistic Hamiltonian
using orbitals calculated in a scaled Thomas-Fermi-Dirac Amaldi
(TFDA) potential. The scaling parameters for this potential
($\lambda_{l}$) have been obtained by a self-consistent energy
minimization procedure, in our case on all term energies of the 21
configurations. Relativistic corrections (spin-orbit, mass, Darwin
and one-body) are also introduced in SST.

We perform a comparison of our energy levels and oscillator
strengths to those published by \citet{Bhatia08} and in the
database NIST \citep{NIST12}. This preliminary comparison is
important since the accuracy of the atomic structure (especially
the oscillator strengths) is a prerequisite for the accuracy of
the line broadening results. We present in Table \ref{tab1},
energy levels for the lowest 20 levels belonging to the
configurations 1s$^{2}(2$s$^{2}$,2s2p, 2p$^{2}$,2s3s, 2s3p, 2s3d).
Our energies are compared to the experimental ones
\citep{Edlen83,Edlen85,Khardi94,Lepson03}, to the 27-configuration
model of \citet{Bhatia08} and to the NIST \citep{NIST12} values,
and an excellent agreement (the difference is less than 1 \%) has
been found between the three results showing that our
21-configuration model provides acceptable atomic structure data.
Oscillator strengths for some transitions from the first five
levels to the lowest ten levels (belonging to the configurations
2s$^{2}$, 2s2p and 2p$^{2}$) are presented in Table \ref{tab2} and
compared to the 27-configuration model of \citet{Bhatia08}. The
relative difference between the two results is about 10 \%. We can
conclude from the preceding comparisons that our atomic structure
study is sufficiently accurate to be adopted in the scattering
problem and thus in the line broadening calculations. Collision
strengths for the same transitions as for the oscillator strengths
are presented in Table \ref{tab2}. Comparison has been made with
the 27-configuration results of \citet{Bhatia08} and an overall
reasonable agreement has been found between the two results. In
some cases, notable differences appear especially for the energy
180 Ry. We can note the case of the transition $1-5$ (2s$^{2}$
$^{1}$S$_{0}-$ 2s2p $^{1}$P$^{\rm o}_{0}$) which is an optical
allowed transition, and it is shown in \citet{Elabidi12} that for
such transitions, whose energy difference $\Delta E$ is very
small, collision strengths can not converge at low total angular
momentum $J^{T}$ especially at high electron energies.

In our line broadening calculations (Eq. \ref{integw}), we use the
imaginary and the real parts of the scattering matrices and these
parameters are related to the corresponding collision strengths.
Consequently, the accuracy of collision strengths presented in
Table \ref{tab2} is very important for the accuracy of our line
broadening data.

\subsection{Line broadening data}
We present in Table \ref{tab3} widths of the four lines: 2s$^{2}$
$^{1}$S$_{0}-$ 2s$n$p $^{1}$P$^{\rm o}_{1}$ where $2\leq n \leq 5$
involving the ground level, and in Table \ref{tab3} the two lines
2s2p $^{1}$P$^{\rm o}_{1}-$2s3s $^{1}$S$_{0}$ and 2s2p
$^{3}$P$^{\rm o}_{0}-$2s3s $^{3}$S$_{1}$ involving excited levels.
Calculations are based on the quantum mechanical (Q) and the
semiclassical perturbation (SCP) formalisms. We note that for all
these transitions, the SCP results are overestimated compared to
the quantum ones, and the average relative difference is about 70
\%. We note also that, except for the resonance line, the two
results Q and SCP become close to each other with the increase of
the principal quantum number $n$. Table \ref{tab4} shows that for
transitions that do not involve the ground level, the SCP results
are no longer higher than the quantum ones. We found also that the
disagreement between the two results is less for these
transitions: the quantum results are about 33 \% higher than the
SCP ones.

To explain at least a part of the previous behaviour of line
widths with the principal quantum number $n$, we present in Table
\ref{tab5} the contributions of strong collisions and those of the
quadrupolar potential for all the considered transitions. We note
that, in general (except for the resonance line), the
contributions of strong collisions and those of the quadrupolar
potential are important for transitions involving levels with low
principal quantum number $n$. Inelastic collisions due to more
distant collisions are quite negligible. We found also that when
the contributions of strong and close collisions, and thus the
contributions of the elastic collisions due to the quadrupolar
potential is dominant, the disagreement between the SCP and the
quantum results is important. For example, for transitions between
excited levels (the two last ones in Table \ref{tab5}), the
relative difference between the SCP and the quantum results are
about 25 \%. In these cases, the strong collisions and the
quadrupolar potential have the lowest contributions (respectively
35 \% and 56 \%) compared to the four other transitions. This
behaviour can be explained by the use of the hydrogenic model for
the atomic structure in the SCP formalism to evaluate the
quadrupolar potential. It is known that this approach
overestimates the corresponding contributions to line widths.

\section{Conclusion}
We have calculated in this work, atomic structure data (energy
levels and oscillator strengths), collision strengths and electron
impact broadening for Ar XV. To check their accuracy, comparisons
of our level energies with the experimental
\citep{Edlen83,Edlen85,Khardi94,Lepson03} and with the theoretical
\citep{Bhatia08, NIST12} results have been performed and a
relative difference of about 1 \% has been found. Our oscillator
strengths have been compared to those of \citet{Bhatia08}, and we
found that the two results agree within 10 \%. For collision
strengths, an overall agreement has been found between our results
and those of \citet{Bhatia08}. This shows firstly that we can
trust our preliminary data and that they can be used with
confidence in our line broadening calculations. For line
broadening, several important results can be derived from our
study. Firstly, we find that the disagreement between the
semiclassical and the quantum results is important when the
contributions of strong and close collisions, and thus the
contributions of the elastic collisions due to the quadrupolar
potential are dominant. In these cases, the semiclassical results
are always higher than the quantum ones. Secondly, we remark that
the contributions of such elastic collisions are important for
transitions involving levels with low principal quantum numbers
$n$ (except for the resonance line). Another point is that for
transitions that do not involve the ground level (for which the
contributions of strong collisions are the smallest), the SCP
results are no longer higher than the quantum ones. Finally, we
can explain the overestimation of the semiclassical line widths
compared to the quantum ones by the fact that the semiclassical
formalism uses the hydrogenic approximation to evaluate the
quadrupolar potential. Extensive works on the strong collision
contributions to the line widths and their behaviour with the
ionization stages along some isoelectronic sequences will be
welcome to investigate their effects on lines broadening and to
study their evaluation in the semiclassical formalism.

\section*{Acknowledgments}
This work has been supported by the Tunisian research unit
05/UR/12-04. It is also a part of the project 176002 "Influence of
collisional processes on astrophysical plasma line shapes"
supported by the Ministry of Education, Science and Technological
Development of Serbia.

\clearpage

\begin{table}
\caption{Our energies in cm$^{-1}$ (Present) for the lowest 20
levels of Ar XV compared to other results. Exp: experimental
energies in \citet{Edlen83,Edlen85,Khardi94,Lepson03} and taken
from \citet{Bhatia08}. NIST: energies from the database NIST
\citep{NIST12}, Bhatia08: calculated energies with a
27-configuration model \citep{Bhatia08}. $i$ labels the 20 levels.
The NIST energies of the two levels 18 and 19 (designed by
asterisks) are inverted compared to all the other results.}
\begin{tabular}{ccccccc}
\hline
\multicolumn{3}{c}{Level designation} &  & &  & \\
$i$ & Conf. & Level & Present & Exp. & NIST & Bhatia08 \\
\hline
1 & 1s$^{2}2$s$^{2}$ & $^{1}$S$_{0}$ & 0 & 0 &  & 0 \\
2 & 1s$^{2}$2s2p & $^{3}$P$_{0}^{\mathrm{o}}$ & 228727 & 228674 &228684 & 229202 \\
3 & 1s$^{2}$2s2p & $^{3}$P$_{1}^{\mathrm{o}}$ & 236470 & 235863 &235860.2 &236662 \\
4 & 1s$^{2}$2s2p & $^{3}$P$_{2}^{\mathrm{o}}$ & 253842 & 252683 &252679.6 &254115 \\
5 & 1s$^{2}$2s2p & $^{1}$P$_{1}^{\mathrm{o}}$ & 459530 & 452212 &452182 &459911 \\
6 & 1s$^{2}$2p$^{2}$ & $^{3}$P$_{0}$ & 608399 & 604961 & 604917 & 609224 \\
7 & 1s$^{2}$2p$^{2}$ & $^{3}$P$_{1}$ & 618807 & 615128 & 615140 & 619718 \\
8 & 1s$^{2}$2p$^{2}$ & $^{3}$P$_{2}$ & 633295 & 628292 & 628308 & 633409 \\
9 & 1s$^{2}$2p$^{2}$ & $^{1}$D$_{2}$ & 698851 & 689621 &  & 699392 \\
10 & 1s$^{2}$2p$^{2}$ & $^{1}$S$_{0}$ & 854805 & 840612 & 840620 & 855441 \\
11 & 1s$^{2}$2s3s & $^{3}$S$_{1}$ & 3938369 & 3935000 &  & 3938375 \\
12 & 1s$^{2}$2s3s & $^{1}$S$_{0}$ & 3983232 & 3980000 & 3980760 & 3981941 \\
13 & 1s$^{2}$2s3p & $^{3}$P$_{1}^{\mathrm{o}}$ & 4044723 & 4042037&  &4044306 \\
14 & 1s$^{2}$2s3p & $^{3}$P$_{0}^{\mathrm{o}}$ & 4046486 & 0 &  & 4045888 \\
15 & 1s$^{2}$2s3p & $^{1}$P$_{1}^{\mathrm{o}}$ & 4051014 & 4042600& 4042040& 4050223 \\
16 & 1s$^{2}$2s3p & $^{3}$P$_{2}^{\mathrm{o}}$ & 4053511 & 4050500&  &4052584 \\
17 & 1s$^{2}$2s3d & $^{3}$D$_{1}$ & 4111931 & 0 & 4106160 & 4110053 \\
18 & 1s$^{2}$2s3d & $^{3}$D$_{2}$ & 4112940 & 0 & 4113330* & 4111049 \\
19 & 1s$^{2}$2s3d & $^{3}$D$_{3}$ & 4114464 & 4110000 & 4109660* &4112559\\
20 & 1s$^{2}$2s3d & $^{1}$D$_{2}$ & 4158547 & 4150000 & 4149860 & 4155932\label{tab1} \\
\hline
\end{tabular}
\end{table}
\clearpage

\begin{table}
\caption{Weighted oscillator strengths $g*f$ and collision
strengths $\Omega$ for transitions from the lowest five levels to
the lowest 10 ones. Present: the present results, Bhatia08:
calculated values from \citet{Bhatia08} with the 27-configuration
model.}
\begin{tabular}{ccccccc}
\hline
Transition & \multicolumn{2}{c}{Oscillator strengths ($g*f$)}& \multicolumn{2}{c}{$ \Omega $ (10 Ry)} & \multicolumn{2}{c}{$\Omega $ (180 Ry)} \\
$i-j$ & Present & Bhatia08 & Present & Bhatia08 & Present & Bhatia08 \\
\hline
$1-2$ &  &  & 2.898E$-$03 & 3.064E$-$03 & 3.530E$-$04 & 3.596E$-$04 \\
$1-3$ & 2.290E$-$04 & 1.203E$-$04 & 8.693E$-$03 & 1.146E$-$02 &4.256E$-$03& 4.305E$-$03 \\
$1-4$ &  &  & 1.449E$-$02 & 1.507E$-$02 & 1.735E$-$03 & 1.765E$-$03 \\
$1-5$ & 2.090E$-$01 & 2.093E$-$01 & 6.753E$-$01 & 6.682E$-$01& 7.623E$-$01 & 1.197E$+$00 \\
$1-6$ &  &  & 1.510E$-$04 & 1.595E$-$04 & 2.600E$-$05 & 2.766E$-$05 \\
$1-7$ &  &  & 4.520E$-$04 & 3.878E$-$04 & 2.100E$-$05 & 2.207E$-$05 \\
$1-8$ &  &  & 7.540E$-$04 & 8.283E$-$04 & 2.150E$-$04 & 2.274E$-$04 \\
$1-9$ &  &  & 3.374E$-$03 & 4.802E$-$03 & 4.585E$-$03 & 4.848E$-$03 \\
$1-10$ &  &  & 1.317E$-$03 & 1.310E$-$03 & 1.021E$-$03 & 1.192E$-$03 \\
$2-3$ &  &  & 3.131E$-$02 & 3.289E$-$02 & 3.066E$-$03 & 3.085E$-$03 \\
$2-4$ &  &  & 2.220E$-$02 & 2.172E$-$02 & 1.776E$-$02 & 1.764E$-$02 \\
$2-5$ &  &  & 7.731E$-$03 & 7.963E$-$03 & 5.970E$-$04 & 6.016E$-$04 \\
$2-6$ &  &  & 1.771E$-$03 & 1.988E$-$03 & 2.140E$-$04 & 2.170E$-$04 \\
$2-7$ & 8.158E$-$02 & 8.167E$-$02 & 3.286E$-$01 & 3.192E$-$01 &3.536E$-$01& 5.789E$-$01 \\
$2-8$ &  &  & 2.214E$-$03 & 3.664E$-$03 & 3.900E$-$04 & 4.094E$-$04 \\
$2-9$ &  &  & 4.695E$-$03 & 3.662E$-$03 & 3.850E$-$04 & 3.929E$-$04 \\
$2-10$ &  &  & 5.490E$-$04 & 4.368E$-$04 & 3.600E$-$05 & 3.616E$-$05 \\
$3-4$ &  &  & 8.910E$-$02 & 8.868E$-$02 & 4.333E$-$02 & 4.340E$-$02 \\
$3-5$ &  &  & 2.319E$-$02 & 2.417E$-$02 & 2.074E$-$03 & 2.091E$-$03 \\
$3-6$ & 7.775E$-$02 & 7.792E$-$02 & 3.286E$-$01 & 3.242E$-$01 &5.866E$-$01& 5.880E$-$01 \\
$3-7$ & 5.972E$-$02 & 5.984E$-$02 & 2.534E$-$01 & 2.466E$-$01& 2.655E$-$01 & 4.361E$-$01 \\
$3-8$ & 1.041E$-$01 & 1.041E$-$01 & 4.158E$-$01 & 4.025E$-$01& 4.409E$-$01 & 7.225E$-$01 \\
$3-9$ & 7.344E$-$04 & 7.387E$-$04 & 1.409E$-$02 & 1.574E$-$02 &4.123E$-$03& 5.680E$-$03 \\
$3-10$ & 6.809E$-$05 & 6.917E$-$05 & 1.646E$-$03 & 1.655E$-$03 &2.950E$-$04& 3.828E$-$04 \\
$4-5$ &  &  & 3.866E$-$02 & 4.095E$-$02 & 3.162E$-$03 & 3.181E$-$03 \\
$4-6$ &  &  & 2.214E$-$03 & 2.015E$-$03 & 2.150E$-$04 & 2.326E$-$04 \\
$4-7$ & 9.494E$-$02 & 8.517E$-$02 & 4.158E$-$01 & 4.095E$-$01& 4.432E$-$01 & 7.369E$-$01 \\
$4-8$ & 2.871E$-$01 & 2.871E$-$02& 1.245E$+$00 & 1.164E$+$00& 1.281E$+$00 & 2.116E$+$00 \\
$4-9$ & 1.170E$-$02 & 1.168E$-$02 & 2.348E$-$03 & 6.612E$-$02& 4.662E$-$02 & 7.232E$-$02 \\
$4-10$ &  &  & 2.743E$-$03 & 3.441E$-$03 & 3.060E$-$04 & 3.103E$-$04 \\
$5-6$ & 2.182E$-$04 & 2.214E$-$04 & 2.790E$-$03 & 5.106E$-$03& 2.934E$-$03 & 5.598E$-$03 \\
$5-7$ & 6.686E$-$05 & 6.663E$-$05 & 8.371E$-$03 & 9.420E$-$03& 1.693E$-$03 & 2.430E$-$03 \\
$5-8$ & 4.230E$-$03 & 4.198E$-$03 & 1.395E$-$02 & 6.147E$-$02& 4.492E$-$02 & 8.579E$-$02 \\
$5-9$ & 2.363E$-$01 & 2.356E$-$01 & 1.776E$+$00 & 1.660E$+$00& 1.697E$+$00 & 3.127E$+$00 \\
$5-10$ & 1.485E$-$01 & 1.490E$-$01 & 5.802E$-$01 &5.506E$-$01& 6.374E$ -$01 & 1.036E$+$00 \label{tab2} \\
\hline
\end{tabular}
\end{table}

\clearpage

\begin{table}
\caption{Present quantum (Q) and semiclassical (SCP) line widths
of some Ar XV transitions involving the ground level.}
\begin{tabular}{ccccc}
\hline
Transition & T($10^{5}$ K) & $Q$ ($10^{-3}$\AA ) & $SCP$ ($10^{-3}$\AA) & $\frac{SCP}{Q}$ \\
\hline
$1s^{2}2s^{2}$$^{1}\mathrm{S}_{0}-2s2p$ $^{1}\mathrm{P}_{1}^{\mathrm{0}}$ &5 & 8.550 & 14.3 & 1.67 \\
$\lambda =221.15$ \AA  & 7.5 & 7.660 & 11.7 & 1.53 \\
& 10 & 7.040 & 10.2 & 1.45 \\
& 20 & 5.620 & 7.46 & 1.33 \\
$1s^{2}2s^{2}$ $^{1}\mathrm{S}_{0}-2s3p$ $^{1}\mathrm{P}_{1}^{\mathrm{0}}$ &5 & 0.242 & 0.455 & 1.88 \\
$\lambda =24.7$ \AA  & 7.5 & 0.209 & 0.373 & 1.78 \\
& 10 & 0.188 & 0.325 & 1.73 \\
& 20 & 0.141 & 0.235 & 1.67 \\
$1s^{2}2s^{2}$ $^{1}\mathrm{S}_{0}-2s4p$ $^{1}\mathrm{P}_{1}^{\mathrm{0}}$ &5 & 0.498 & 0.758 & 1.52 \\
$\lambda =18.8$ \AA  & 7.5 & 0.398 & 0.634 & 1.59 \\
& 10 & 0.338 & 0.560 & 1.66 \\
& 20 & 0.224 & 0.422 & 1.88 \\ $1s^{2}2s^{2}$ $^{1}\mathrm{S}_{0}-2s5p$ $^{1}\mathrm{P}_{1}^{\mathrm{0}}$ &5 & 0.884 & 1.37 & 1.55 \\
$\lambda =16.95$ \AA  & 7.5 & 0.693 & 1.17 & 1.69 \\
& 10 & 0.580 & 1.04 & 1.79 \\
& 20 & 0.370 & 0.805 & 2.18 \label{tab3} \\
\hline
\end{tabular}
\end{table}

\begin{table}
\caption{Present quantum (Q) and semiclassical (SCP) line widths
of two Ar XV transitions between excited levels.}
\begin{tabular}{ccccc}
\hline
Transition & T($10^{5}$ K) & $Q$ ($10^{-3}$\AA ) & $SCP$ ($10^{-3}$\AA ) & $\frac{SCP}{Q}$ \\
\hline
$1s^{2}2s2p$ $^{3}\mathrm{P}_{0}^{\mathrm{0}}-2s3s$$^{3}\mathrm{S}_{1}$ & 5& 0.331 & 0.242 & 0.73 \\
$\lambda =27.0$ \AA  & 7.5 & 0.273 & 0.202 & 0.74 \\
& 10 & 0.236 & 0.177 & 0.75 \\
& 20 & 0.164 & 0.133 & 0.81 \\
$1s^{2}2s2p$$^{1}\mathrm{P}_{1}^{\mathrm{0}}-2s3s$ $^{1}\mathrm{S}_{0}$ & 5& 0.363 & 0.280 & 0.77 \\
$\lambda =28.4$ \AA  & 7.5 & 0.309 & 0.232 & 0.75 \\
& 10 & 0.273 & 0.204 & 0.75 \\
& 20 & 0.197 & 0.152 & 0.77 \label{tab4} \\
\hline
\end{tabular}
\end{table}

\begin{table}
\caption{Strong collisions (strong) and quadrupolar potential
(quad) contributions to line widths.}
\begin{tabular}{cccc}
\hline
& strong/total$(\%)$ & quad/total$(\%)$ & $\frac{\left\vert SCP-Q\right\vert }{Q}(\%)$ \\
\hline
$2s^{2}$$^{1}\mathrm{S}_{0}-2s2p$ $^{1}\mathrm{P}_{1}^{\mathrm{0}}$ & $42$& $65$ & $67$ \\
$2s^{2}$ $^{1}\mathrm{S}_{0}-2s3p$$^{1}\mathrm{P}_{1}^{\mathrm{0}}$ & $56$& $88$ & $88$ \\
$2s^{2}$ $^{1}\mathrm{S}_{0}-2s4p$$^{1}\mathrm{P}_{1}^{\mathrm{0}}$ & $45$& $71$ & $52$ \\
$2s^{2}$ $^{1}\mathrm{S}_{0}-2s5p$$^{1}\mathrm{P}_{1}^{\mathrm{0}}$ & $38$& $60$ & $55$ \\
$2s2p$ $^{3}\mathrm{P}_{0}^{\mathrm{0}}-2s3s$ $^{3}\mathrm{S}_{1}$& $35$ & $56$ & $27$ \\
$2s2p$ $^{1}\mathrm{P}_{1}^{\mathrm{0}}-2s3s$ $^{1}\mathrm{S}_{0}$& $36$ & $ 56$ & $23$ \label{tab5}\\
\hline
\end{tabular}
\end{table}


\end{document}